# Performance of New High-Precision Muon Tracking Detectors for the ATLAS Experiment

Oliver Kortner, Hubert Kroha*, Korbinian Schmidt-Sommerfeld, Eric Takasugi

Max-Planck-Institut für Physik, Föhringer Ring 6, D-80805 Munich, Germany

*Abstract*–The goals of the ongoing and planned ATLAS muon detector upgrades are to increase the acceptance for precision muon momentum measurement and triggering and to improve the rate capability of the muon chambers in the high-background regions corresponding to the increasing LHC luminosity. Small-diameter Muon Drift Tube (sMDT) chambers have been developed for these purposes. With half the drift-tube diameter of the current ATLAS Muon Drift Tube (MDT) chambers with 30 mm drift tube diameter and otherwise unchanged operating parameters, the sMDT chambers share all the advantages of the MDTs, but have an about an order of magnitude higher rate capability and can be installed in detector regions where MDT chambers do not fit in. The construction of twelve chambers for the feet regions of the ATLAS detector has been completed for the installation in the winter shutdown 2016/17 of the Large Hadron Collider. The purpose of this upgrade of the ATLAS muon spectrometer is to increase the acceptance for three-point muon track measurement which substantially improves the muon momentum resolution in the regions concerned.

## I. CHAMBER CONSTRUCTION

The design of the new sMDT chambers [1,2] is shown in Fig. 1. The 4500 drift tubes for these chambers have been assembled and tested between October 2014 and end of 2015 using a semi-automated wiring station in a temperature controlled clean room. Each drift tube had to pass stringent tests of wire tension, gas tightness at the operating pressure of 3 bar and leakage current under the operating voltage of 2730 V with Ar:$CO_2$ (93:7) drift gas before assembly in a chamber.

Here we report about the construction and test of 12 new sMDT chambers which started in October 2015 and has been completed in November 2016. The drift tube design (see Fig.2) and the chamber construction procedures (see Fig. 3) have been optimized for mass production while providing very high mechanical accuracy. An sMDT chamber can be assembled within one working day independent of the number of tube layers while the construction of the MDT chambers [3] required one day per tube layer. The external reference surfaces of the brass inserts of the endplugs of the drift tubes, which are concentric with the internal wire locators with high precision, are inserted into fitting holes in the chamber assembly jigs. The jigs define the wire grid at each chamber end and are stacked together with the tube layers (see Fig.2).

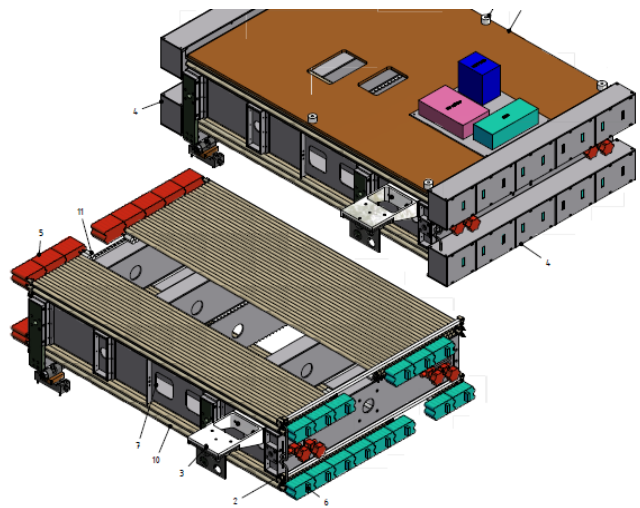

Fig. 1: Design of the new sMDT chambers for the middle barrel layer of the ATLAS muon spectrometer (BMG chambers). The chambers consist of about 350 drift tubes of 15 mm diameter and 1.1 m length which are arranged in two quadruple layers separated by an aluminum space frame. The extension brackets for precise mounting of the optical alignment monitoring sensors with respect to the sense wires are shown. The cutouts in the tube layers allow for the passage of the light rays connecting the chamber layers of the muon spectrometer for alignment monitoring purposes.

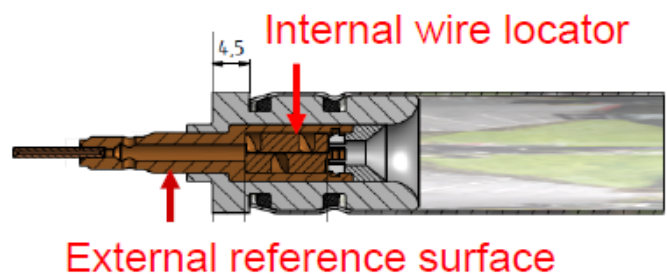

Fig. 2: Design of the endplugs of the sMDT drift tubes. The endplug consists of an injection molded plastic insulator (grey) with a brass insert (brown) which holds the spiral-shaped wire locator on the inside of the tube positioning the sense wire with micron accuracy with respect to a cylindrical reference surface on the outside of the tube. The reference surfaces of the endplugs are used for positioning of the tubes during chamber assembly (see Fig. 3) and for measuring the relative positions of all sense wires in an assembled chamber (see Fig. 4).

* Corresponding author (kroha@mppmu.mpg.de).

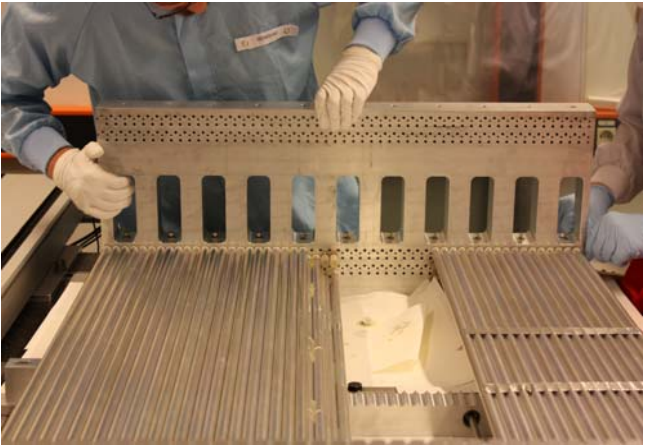

Fig. 3: Positioning of the drift tubes with the reference surfaces of the endplugs (see Fig.1) in hole grids predefined by precision jigs at each chamber end which have been machined with 5 µm accuracy and are stacked together with the tube layers.

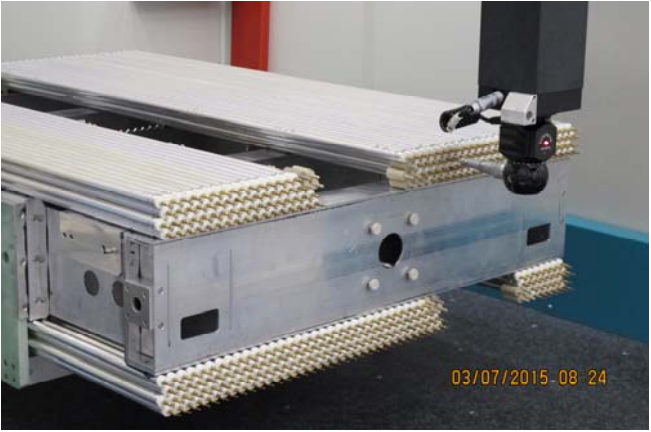

Fig. 4: The external reference surfaces of the endplugs allow for measurement of the relative wire positions of all drift tubes in an assembled chamber using a commercial coordinate measuring machine.

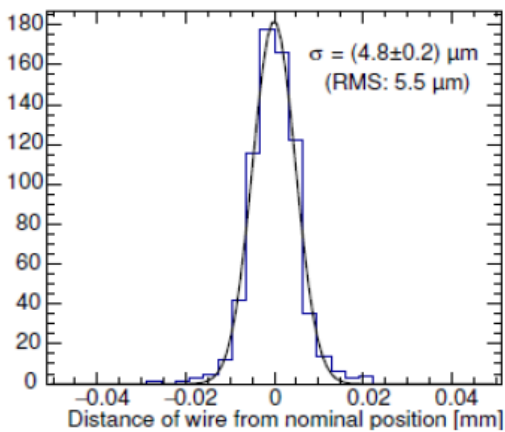

Fig. 5: Distribution of the radial displacements of the sense wire positions from the nominal wire grid at both ends of an sMDT chamber measured with a coordinate measuring machine (see Fig. 4).

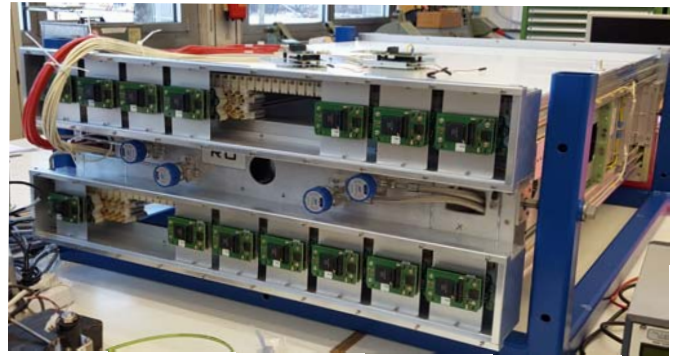

Fig. 6: A completed sMDT chamber for installation in the feet regions of the ATLAS barrel muon spectrometer in the LHC winter shutdown 2016/17 with mounted readout electronics boards ready for comic ray test.

## II. ASSEMBLY ACCURACY

A wire positioning accuracy of better than 20 µm (rms) is required for the sMDT chambers like for the MDT chambers before. In the construction of the new sMDT chambers, a four times better wire positioning accuracy of 5 µm (rms) has been achieved (see Fig. 5) which is unprecedented. The relative wire positions of the drift tubes in each chambers have been measured using a commercial coordinate measuring machine (see Fig. 4).

After the wire position measurement (which includes also the measurement of the alignment sensor mounting brackets with respect to the wire grid), the gas distribution system and the high-voltage distribution and readout electronics boards are mounted (see Fig. 6). Afterwards the chambers have been tested again with respect to gas leak rate, leakage currents and electronics noise as well as with cosmic ray muons before and after shipment to CERN.